\documentclass[11pt,a4paper]{article}
\usepackage[margin=2.5cm]{geometry}
\usepackage{amsmath,amssymb}
\usepackage{bm}
\usepackage{graphicx}
\usepackage{hyperref}
\title{Entropy Has No Direction: A Mirror-State Paradox Against Universal Monotonic Entropy Increase\\
and a First-Principles Proof that Constraints Reshape the Entropy Distribution $P_{\infty}(S;\lambda)$\\[0.5em]
\large An Adventure Toward Energy Freedom and the Rejection of Universal Heat Death}
\author{Ting Peng\\
  \texttt{t.peng@ieee.org}\\
  ORCID: \href{https://orcid.org/0009-0001-9059-2278}{0009-0001-9059-2278}\\
  Key Laboratory for Special Area Highway Engineering of Ministry of Education,\\
  Chang'an University, Xi'an 710064, China}
\date{\today}

\begin{document}
\maketitle

\noindent\footnotesize The \LaTeX\ source of this manuscript, together with a live FAQ, is available at \url{https://github.com/tpeng1977/entropy}.

\begin{abstract}
We revisit textbook claims that entropy must increase and show that, under
time-reversal invariant microscopic dynamics, no universal trajectory-wise or
statistical assertion that the coarse-grained entropy $S(t)$ is non-decreasing
can hold. The core is a mirror-state construction: for any microstate $A$ one
constructs its time-reversed partner $B$ (momenta inverted); requiring $S(t)$
to be non-decreasing for both $A$ and $B$ forces every time to be a local
minimum of $S$ and hence makes $S(t)$ constant along the trajectory. The
consistent picture is that entropy is a stochastic variable described by a
probability distribution $P(S)$ whose shape depends on constraints and boundary
conditions; entropy-based regularities are emergent summaries of
constraint-dependent microscopic dynamics, and in practice it is constraints
and boundaries---not entropy itself---that one manipulates to achieve mixing,
separation, or self-organization.
Working with Boltzmann (coarse-grained) entropy on the energy shell, we then
derive from first principles how constraints reshape the long-time entropy
distribution $P_{\infty}(S;\lambda)$ by altering the invariant measure through
changes in the Hamiltonian and/or the accessible phase space. In the
microcanonical setting we obtain a sharp criterion: the \emph{only} way
$P_{\infty}^{(E)}(S;\lambda)$ can remain the same up to translation is when all
accessible macrostate volumes are scaled by a common factor; otherwise the
distribution changes structurally. We connect this framework to experiments on
asymmetric nanopores and molecular gates, to macroscopic examples from civil
engineering (windbreak forests, dikes, vortex suppression, traffic-flow
control), and to natural phenomena such as lightning guided to lightning rods,
snowflake and mineral-veil growth, and the sudden crystallisation of supercooled
water. In all these cases, constraints and dynamics together determine the
long-time entropy distribution and can make ordered, low-entropy macrostates
statistically favoured compared with the unconstrained case.
\end{abstract}

\section{Introduction}

To enable safe, comfortable, and rapid travel, we have overcome geological and
environmental obstacles, spanned natural chasms, and turned the impossible into
the possible---building highways, railways, and airports on a vast scale. These
achievements stand as a cornerstone of civilisation. They rest on a single
principle: constraints reshape what is accessible, and thus what becomes
typical. This paper makes that principle precise and shows that it extends from
macroscopic engineering to the foundations of thermodynamics.

In an age when robotics and artificial intelligence have become commonplace,
many past myths and ``impossibilities'' have become reality. If a perpetual
motion machine of the second kind can be realised, we may achieve energy
freedom; if the universe is destined for heat death, then all our efforts will
come to nothing. Driven by the desire for energy freedom and the refusal to
accept the heat-death thesis, we take logic and mathematics as our tools and
embark on this technical exploration. That exploration begins with the law we
are to re-examine.

The ``Second Law'' is often presented in two forms.
\emph{Strict (deterministic) form}: for an isolated system, entropy does not
decrease along \emph{every} trajectory, i.e. $S(t+\delta t)\ge S(t)$ for all $t$
and all microstates (or $\mathrm{d}S/\mathrm{d}t\ge 0$ wherever defined).
\emph{Statistical form}: for an appropriate ensemble or limit, entropy increase is
``overwhelmingly probable'' or the ensemble-average entropy change is nonnegative.

This paper has a single scope: we address these statements only insofar as they are
claimed to be \emph{universal} consequences of time-reversal invariant microscopic
physics. We show that such universal monotonicity claims are logically incompatible
with time-reversal symmetry. The consistent replacement is not to end thermodynamics,
but to replace ``entropy has a direction'' with: \textbf{entropy has no direction;
it is described by a probability distribution}. Moreover, \textbf{constraints and
boundary conditions reshape this distribution}. We prove from first principles that
constraints reshape the long-time entropy distribution $P_{\infty}(S;\lambda)$.
That proof was motivated in part by the author's earlier simulation work
showing that geometry can challenge conventional entropy regimes in nanofluidic
cascades~\cite{peng2026geometrychallengesentropyregimedependentrectification}.
We then discuss recent experimental work that validates this framework.

\section{Preliminaries and assumptions}
\label{sec:prelim}

\subsection{Microscopic dynamics and time reversal}

Let $\Gamma(t)$ denote the phase-space microstate of an isolated system evolving
under time-reversal invariant microscopic dynamics (Hamiltonian dynamics being the
canonical example). Let $\mathcal{T}$ denote time reversal, which (for classical
mechanics) acts by reversing momenta while leaving positions unchanged.

Time-reversal invariance means: if $\Gamma_A(t)$ is a solution, then
$\Gamma_B(t):=\mathcal{T}\Gamma_A(2t_0-t)$ is also a solution. In particular, the
forward-time evolution of the mirror state $\Gamma_B(t_0)=\mathcal{T}\Gamma_A(t_0)$
replays the past of $\Gamma_A$ in reverse.

\subsection{Entropy as a coarse-grained state function}

We use Boltzmann (coarse-grained) entropy: fix a time-reversal symmetric
coarse-graining (partition) of phase space into macrostates (cells) $\{C_m\}$.
Assign to any $\Gamma\in C_m$ an instantaneous entropy $S(\Gamma)=S_m$.

For mathematical cleanliness (finite volumes), we adopt the energy-shell form used
in statistical mechanics: at fixed energy $E$ define the accessible macrostate
volume $W_m^{(E)}$ on the energy shell and set $S_m^{(E)}=k_B\ln(W_m^{(E)}/W_0)$,
with an arbitrary reference volume $W_0$ (shifting entropy by a constant does not
affect any conclusions). This is the same entropy notion used below to derive
$P_{\infty}(S;\lambda)$.

We assume the coarse-graining is time-reversal symmetric, so $S(\mathcal{T}\Gamma)=S(\Gamma)$.

\subsection{Continuity (minimal regularity)}
\label{sec:continuity}

The mirror-state paradox below is clearest when $S(t):=S(\Gamma(t))$ is continuous
in time along trajectories (standard for smooth coarse-graining, or for coarse
variables defining macrostates). This is the only regularity used to turn ``local
minimum everywhere'' into ``constant everywhere''.

\paragraph{Remark (discrete coarse-graining).}
If the coarse-graining is strictly discrete, then $S(t)$ may be piecewise constant
with jumps. The paradox still goes through for a universal monotonicity claim that
is asserted for \emph{arbitrarily small} $\delta t>0$: Eq.~\eqref{eq:two_sided_min}
forces $S(t_0)$ to be simultaneously a right- and left-minimum at every $t_0$,
which rules out any jump up or down and again implies that $S(t)$ is constant.

\subsection{What is meant by $P_{\infty}(S;\lambda)$}
\label{sec:Pinf_meaning}

In the second part of the paper, $P_{\infty}(S;\lambda)$ denotes a \emph{long-time}
entropy distribution induced by an invariant measure. Two standard routes make
this precise:
(i) assume the dynamics are ergodic/mixing on the relevant invariant set so that
time averages (and long-time distributions) are independent of the initial
microstate up to measure-zero exceptions; or
(ii) consider an ensemble whose initial microstates are drawn from the invariant
measure (microcanonical or canonical), in which case $P_{\infty}$ is immediate.
Our formulas below are statements about the invariant measures themselves and the
entropy distributions induced by them; the above routes justify interpreting these
as long-time distributions.

\section{The mirror-state paradox}
\label{sec:paradox}

\subsection{Strict (trajectory-wise) monotonicity is impossible as a universal law}

\textbf{Claim.} A universal law of the form ``for every microstate and every time,
$S(t+\delta t)\ge S(t)$ for all sufficiently small $\delta t>0$'' is incompatible
with time-reversal invariant microscopic dynamics and time-reversal symmetric
coarse-grained entropy.

\textit{Proof (mirror-state construction).}
Fix an arbitrary time $t_0$ on an arbitrary trajectory $\Gamma_A(t)$. Construct the
mirror state at the same time: $\Gamma_B(t_0)=\mathcal{T}\Gamma_A(t_0)$, and let
$\Gamma_B(t)$ be its forward-time evolution. By time-reversal invariance, for any
$\delta t>0$,
\[
\Gamma_B(t_0+\delta t)=\mathcal{T}\Gamma_A(t_0-\delta t).
\]
By time-reversal symmetry of the coarse-graining, $S(\mathcal{T}\Gamma)=S(\Gamma)$,
so
\[
S_B(t_0+\delta t)=S_A(t_0-\delta t),
\qquad
S_B(t_0)=S_A(t_0).
\]
Now apply the universal monotonicity statement to \emph{both} $A$ and $B$:
\[
S_A(t_0+\delta t)\ge S_A(t_0),
\qquad
S_B(t_0+\delta t)\ge S_B(t_0).
\]
The second inequality becomes $S_A(t_0-\delta t)\ge S_A(t_0)$. Hence, for all small
$\delta t>0$,
\begin{equation}
S_A(t_0+\delta t)\ge S_A(t_0)
\quad\text{and}\quad
S_A(t_0-\delta t)\ge S_A(t_0).
\label{eq:two_sided_min}
\end{equation}
Thus $t_0$ is a (two-sided) local minimum of $S_A(t)$.

Because $t_0$ was arbitrary, \emph{every} time is a local minimum. If $S_A(t)$ is
continuous (Sec.~\ref{sec:continuity}), a function for which every point is a local
minimum must be constant. Therefore the universal strict Second Law implies that
entropy is constant on every trajectory, which contradicts the empirical fact that
macroscopic systems exhibit entropy increase and eliminates any entropic arrow of
time. \hfill $\square$

\begin{figure}[t]
\centering
\includegraphics[width=\textwidth]{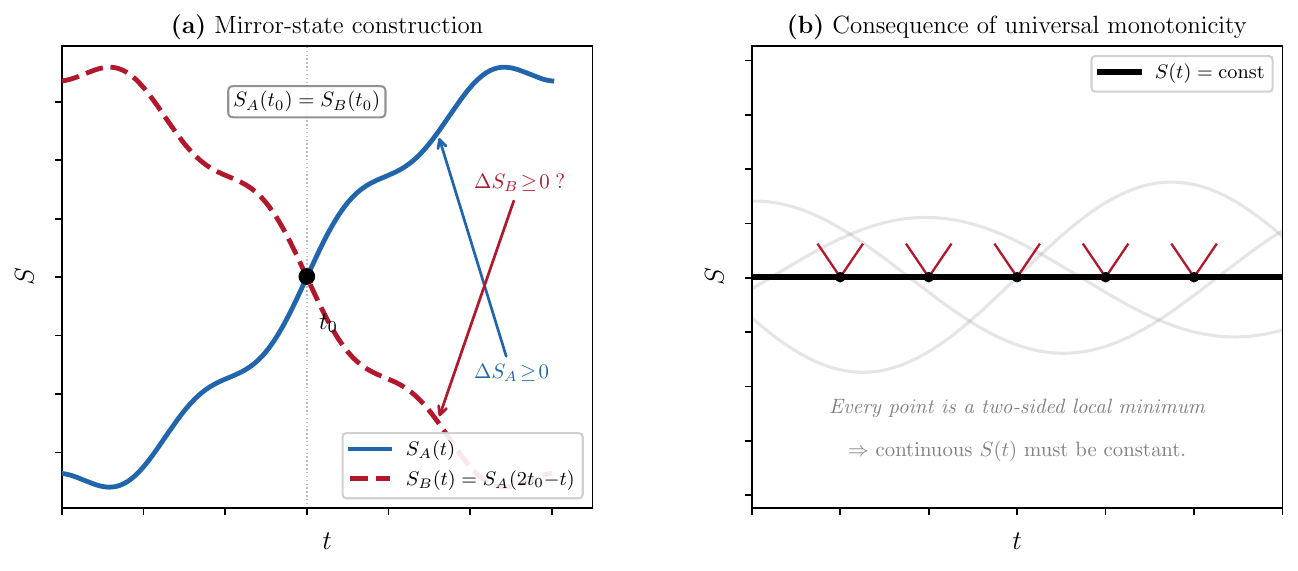}
\caption{The mirror-state paradox.
\textbf{(a)}~For any trajectory $S_A(t)$ (solid blue) and its time-reversed
partner $S_B(t)=S_A(2t_0-t)$ (dashed red), universal monotonicity would require
both $\Delta S_A\ge 0$ and $\Delta S_B\ge 0$ forward from $t_0$.
Because $S_B(t_0+\delta t)=S_A(t_0-\delta t)$, the second condition forces
$S_A$ to be non-decreasing \emph{backward} as well, making $t_0$ a two-sided
local minimum.
\textbf{(b)}~Since $t_0$ is arbitrary, every point must be a local minimum;
for a continuous function this implies $S(t)=\mathrm{const}$, contradicting
observed entropy changes.}
\label{fig:mirror}
\end{figure}

\subsection{Extension: the statistical Second Law cannot be universal}
\label{sec:stat}

The same mirror-state logic refutes the statistical Second Law \emph{when it is
asserted as a universal principle applying to every microstate, including its
mirror}. Suppose one claims, universally, that entropy increase is ``overwhelmingly
probable'' for the forward-time evolution of \emph{any} initial microstate. Apply
this to $A$ at time $t_0$, and also to $B=\mathcal{T}A$ at $t_0$. Since the forward
evolution of $B$ corresponds to the time-reversed past of $A$, ``overwhelmingly
probable increase'' for both implies (with overwhelming probability) the two-sided
inequalities in Eq.~\eqref{eq:two_sided_min}, i.e. that $t_0$ is a local minimum
with overwhelming probability. Choosing $t_0$ arbitrarily destroys any persistent
arrow of time in the same way.

Therefore, the statistical Second Law cannot be a universal state-by-state principle.
Any one-way statement about $\Delta S$ must depend on additional structure: special
initial ensembles, coarse-graining choices, limits, or constraints/boundaries that
select a particular effective description. This observation motivates the corrected
view below.

\section{Corrected view: entropy is a random variable\\with a constraint-dependent PDF}
\label{sec:corrected}

The consistent replacement of ``entropy has a direction'' is:
\textbf{entropy has no direction; it has a probability distribution}.
Write $P_t(S)$ for the (time-dependent) entropy distribution induced by an
ensemble of microstates at time $t$, and $P_{\infty}(S;\lambda)$ for the long-time
distribution under constraint parameters $\lambda$ (geometry, boundary conditions,
static fields, etc.).

At the microscopic level, however, nothing in the dynamics ``sees'' entropy.
Individual particles obey time-reversal invariant equations of motion; their
trajectories and interactions are governed by the Hamiltonian and forces, not by
macroscopic state functions such as entropy. Boltzmann entropy is a coarse-grained,
statistical description of the collective behaviour of many particles---a
macroscopic \emph{appearance} of the underlying time-reversal invariant motion,
obtained only after we partition phase space into macrostates and count accessible
microstates. Constraints cannot change the fundamental microscopic rules; their
core role is to intervene directly in the actual motion of particles by changing
the Hamiltonian $H(\Gamma;\lambda)$ and/or the accessible set
$\mathcal{A}(\lambda)$, thereby restricting where particles can go and how they
move. This, in turn, changes the number of accessible microstates for each
macrostate and thus reshapes the entropy distribution. All changes in entropy are
therefore \emph{statistical summaries} of how constraints redirect microscopic
trajectories. Entropy is a diagnostic, not a causal driver. In practical
engineering terms, what one truly controls and optimises are constraints and
boundary conditions (to achieve, say, separation or mixing), while entropy is
useful as a bookkeeping tool for whether a given design makes certain macrostates
typical or rare. From a logical perspective it is more meaningful to study how
different constraints alter system dynamics and long-time distributions than to
postulate universal laws about ``entropy increase'' detached from the underlying
microphysics.

From this angle, familiar macroscopic ``entropy increase'' phenomena are best
understood as follows. In many textbook and laboratory situations, the initial
state is prepared to be \emph{very special} and low-entropy (gas in one half of
a box, sharp temperature gradients, unmixed components, etc.). Starting from such
atypical initial conditions, the entropy---as a coarse-grained statistic of
particle configurations---does indeed rise rapidly toward the values typical of
the long-time distribution under the given constraints. Once the system has
reached that regime, however, the entropy does not freeze at a single ``equilibrium
value'': microscopic dynamics continue indefinitely and $S(t)$ keeps fluctuating,
with spontaneous increases and decreases both being normal. The role of constraints
is to determine the \emph{shape} of the long-time distribution for $S$ (and for
other observables), not to enforce monotone drift toward a uniquely defined maximum.
There need not be a strict equilibrium state in the sense of a single entropy
value, nor must the formal ``maximum entropy'' compatible with constraints coincide
with the most probable coarse-grained macrostate once dynamical restrictions are
taken seriously.

What fundamentally decides whether a system mixes, separates, or self-organises
into ordered patterns is therefore the constraint-dependent microscopic dynamics
and the resulting invariant measures---not an entropy ``law'' acting as a causal
agent. Accordingly, all results in this paper concerning entropy distributions and
their evolution should be read as probabilistic, ensemble-level characterisations
of typical behaviour under specified constraints, rather than as per-trajectory
guarantees for every single realisation of a physical system.

The remainder of this paper proves the second core claim---\textbf{constraints and
boundaries can change the long-time entropy distribution
$P_{\infty}(S;\lambda)$} in an explicit, first-principles way---and then presents
experimental validation of this framework.

\section{First-principles derivation: constraints $\lambda \to P_{\infty}(S;\lambda)$}
\label{sec:constraint_to_P}

\subsection{Constraints as Hamiltonian/accessible-set modifications}

Let $\lambda$ denote constraint parameters. Constraints enter through the Hamiltonian
\begin{equation}
H(\Gamma;\lambda)=H_0(\Gamma)+V_{\mathrm{c}}(\Gamma;\lambda),
\label{eq:H_constraint}
\end{equation}
and/or through an accessible set $\mathcal{A}(\lambda)$ (hard constraints). Both
views are equivalent: hard walls correspond to $V_{\mathrm{c}}=+\infty$ outside
$\mathcal{A}(\lambda)$.

\subsection{Long-time invariant measures (microcanonical and canonical)}

\paragraph{Microcanonical (isolated).}
At fixed energy $E$, a natural invariant measure is uniform on the accessible energy shell:
\begin{equation}
\rho_{\infty}^{(E)}(\Gamma;\lambda)=
\frac{1}{\Omega(E;\lambda)}\,
\delta\!\bigl(H(\Gamma;\lambda)-E\bigr)\,
\mathbf{1}_{\mathcal{A}(\lambda)}(\Gamma),
\label{eq:rho_micro}
\end{equation}
where the accessible density of states is
\begin{equation}
\Omega(E;\lambda)=
\int \mathrm{d}\Gamma\,
\delta\!\bigl(H(\Gamma;\lambda)-E\bigr)\,
\mathbf{1}_{\mathcal{A}(\lambda)}(\Gamma).
\label{eq:Omega_entropy}
\end{equation}

\paragraph{Canonical (heat bath).}
With a heat bath at temperature $T$ and dynamics obeying fluctuation--dissipation,
the invariant measure is
\begin{equation}
\rho_{\infty}(\Gamma;\lambda)=
\frac{1}{Z(\lambda)}\,e^{-\beta H(\Gamma;\lambda)},
\qquad
Z(\lambda)=\int_{\Gamma\in\mathcal{A}(\lambda)} \mathrm{d}\Gamma\,
e^{-\beta H(\Gamma;\lambda)}.
\label{eq:rho_can}
\end{equation}

\subsection{Boltzmann entropy on the energy shell}

Fix a time-reversal symmetric coarse-graining $\{C_m\}_{m\in\mathcal{M}}$.
Define the accessible macrostate volume on the energy shell:
\begin{equation}
W_m^{(E)}(\lambda)=
\int \mathrm{d}\Gamma\,
\delta\!\bigl(H(\Gamma;\lambda)-E\bigr)\,
\mathbf{1}_{C_m}(\Gamma)\,
\mathbf{1}_{\mathcal{A}(\lambda)}(\Gamma),
\label{eq:WmE_entropy}
\end{equation}
and assign the Boltzmann entropy value
\begin{equation}
S_m^{(E)}(\lambda)=k_B\ln\!\left(\frac{W_m^{(E)}(\lambda)}{W_0}\right).
\label{eq:SmE_entropy}
\end{equation}

\subsection{Closed-form expression for $P_{\infty}(S;\lambda)$ and a sharp change criterion}

\paragraph{Microcanonical.}
Because the microcanonical measure is uniform on the energy shell, the macrostate
probability is a ratio of accessible volumes:
\begin{equation}
\pi_m^{(E)}(\lambda)=\frac{W_m^{(E)}(\lambda)}{\Omega(E;\lambda)},
\qquad
\Omega(E;\lambda)=\sum_{m\in\mathcal{M}}W_m^{(E)}(\lambda).
\label{eq:pi_micro_entropy}
\end{equation}
Therefore the long-time entropy distribution at fixed energy is
\begin{equation}
\boxed{
P_{\infty}^{(E)}(S;\lambda)=
\sum_{m\in\mathcal{M}}
\frac{W_m^{(E)}(\lambda)}{\Omega(E;\lambda)}\,
\delta\!\bigl(S-S_m^{(E)}(\lambda)\bigr).
}
\label{eq:PS_micro_entropy}
\end{equation}

\textbf{Proposition (sharp criterion; ``only translation'' degeneracy).}
Fix $E$ and the coarse-graining. Let $\mathcal{V}(\lambda)$ denote the multiset of
accessible macrostate volumes $\{W_m^{(E)}(\lambda)\}_{m\in\mathcal{M}}$ (counting multiplicity).
\begin{enumerate}
  \item If there exists $c>0$ such that $\mathcal{V}(\lambda_2)=c\,\mathcal{V}(\lambda_1)$
  (i.e. all macrostate volumes scale by a common factor, up to permutation), then
  \[
  P_{\infty}^{(E)}(S;\lambda_2)=P_{\infty}^{(E)}\!\left(S-k_B\ln c\,;\lambda_1\right),
  \]
  i.e. the distribution changes only by translation of the entropy axis.
  \item Otherwise, $P_{\infty}^{(E)}(S;\lambda_2)$ is \emph{not} a translate of
  $P_{\infty}^{(E)}(S;\lambda_1)$; the distribution changes structurally.
\end{enumerate}

\textit{Proof.}
Let $M$ be the macrostate index with $\mathbb{P}(M=m)=\pi_m^{(E)}(\lambda)$ and define
$X_\lambda:=W_M^{(E)}(\lambda)$. Then $S=k_B\ln(X_\lambda/W_0)$ and
Eq.~\eqref{eq:PS_micro_entropy} is exactly the law of $S$.
If $W_m^{(E)}(\lambda_2)=c\,W_{\sigma(m)}^{(E)}(\lambda_1)$ for some permutation $\sigma$, then
$X_{\lambda_2}\overset{d}{=}c\,X_{\lambda_1}$, hence
$S_{\lambda_2}\overset{d}{=}S_{\lambda_1}+k_B\ln c$, which is the claimed translation.
Conversely, if $P_{\infty}^{(E)}(S;\lambda_2)$ were a translate of $P_{\infty}^{(E)}(S;\lambda_1)$ by
$\Delta$, then $X_{\lambda_2}=W_0 e^{S/k_B}$ would be distributed as $e^{\Delta/k_B}X_{\lambda_1}$,
which implies $\mathcal{V}(\lambda_2)=e^{\Delta/k_B}\mathcal{V}(\lambda_1)$ (up to permutation).
\hfill $\square$

\begin{figure}[t]
\centering
\includegraphics[width=\textwidth]{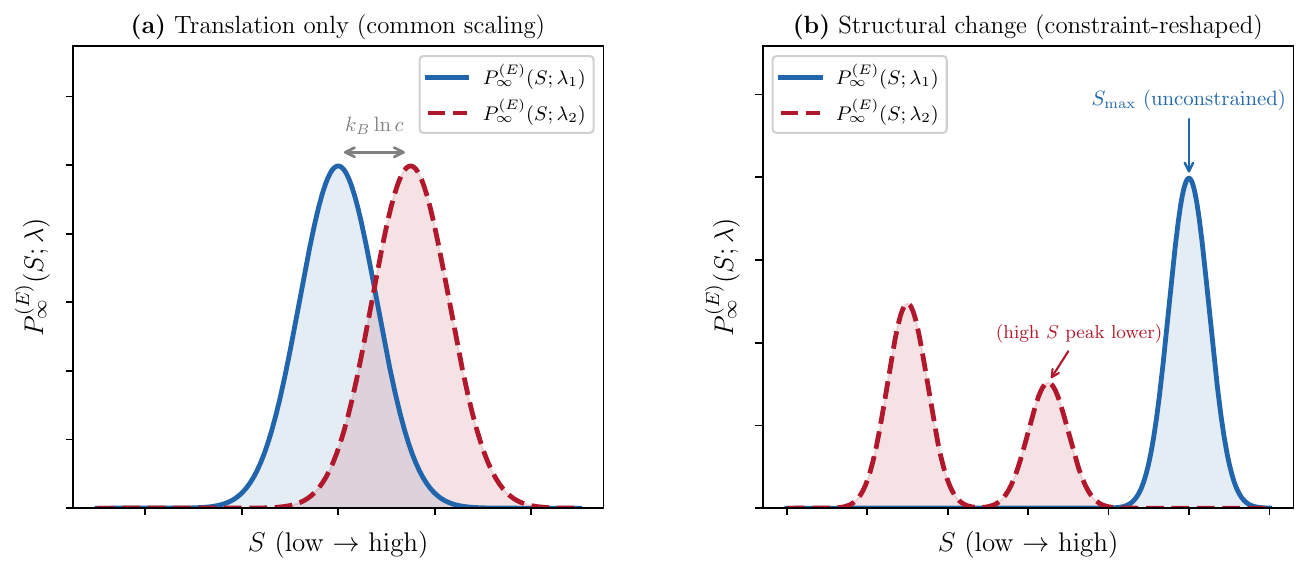}
\caption{How constraints reshape the long-time entropy distribution
$P_{\infty}^{(E)}(S;\lambda)$. Here $P_\infty$ is the \emph{distribution} of
entropy over long time; the system keeps fluctuating and there is no single
equilibrium state. \textbf{(a)}~When all macrostate volumes scale by a common
factor~$c$, the distribution translates along the $S$-axis by $k_B\ln c$
without changing shape (the only degenerate case in the sharp criterion).
\textbf{(b)}~Unconstrained case: entropy is almost entirely concentrated
near the \emph{maximum} $S_{\max}$ (narrow peak at high $S$). Under a
generic constraint the distribution reshapes: significant probability
mass moves to lower $S$, making lower-entropy
macrostates statistically accessible.}
\label{fig:Pinf}
\end{figure}

\paragraph{Canonical as an exact energy mixture.}
The canonical energy density is
\begin{equation}
P_\beta(E;\lambda)=\frac{\Omega(E;\lambda)\,e^{-\beta E}}{Z(\lambda)},
\qquad
Z(\lambda)=\int \mathrm{d}E\;\Omega(E;\lambda)\,e^{-\beta E}.
\label{eq:PE_entropy}
\end{equation}
Since $P_{\infty}^{(E)}(S;\lambda)$ is the entropy distribution conditioned on energy $E$,
the canonical long-time entropy distribution is the exact mixture
\begin{equation}
\boxed{
P_{\infty,\beta}(S;\lambda)=\int \mathrm{d}E\;P_\beta(E;\lambda)\,P_{\infty}^{(E)}(S;\lambda).
}
\label{eq:PS_can_mix_entropy}
\end{equation}

\subsection{Bulletproof qualifier: changing rates is not changing the invariant measure}

The above results concern changes that alter the invariant measure by changing
$\mathcal{A}(\lambda)$ and/or $H(\Gamma;\lambda)$.
By contrast, a purely kinetic modification that only rescales crossing rates or
mixing times \emph{while leaving the invariant measure unchanged} can change the
finite-time distribution $P_t(S)$ but does not change the long-time distribution
$P_{\infty}(S)$.

\section{Experimental validation: asymmetric constraints enable spontaneous low-entropy transitions}
\label{sec:experimental}

Recent experimental work by Qiao and Wang~\cite{qiao2025intrinsicnonequilibriumdistributionlarge}
provides compelling validation of the theoretical framework presented above. Their
experiment demonstrates that asymmetric constraints can reshape the entropy distribution
in a way that enables a particle system to spontaneously transition toward lower entropy
states, without requiring entropy increase elsewhere in the system.

\subsection{The Qiao--Wang experiment}

Qiao and Wang investigated nanoporous carbon electrodes in dilute aqueous CsPiv
solutions where the effective nanopore size ($d_e \approx 1$\,nm) only slightly
exceeds the ion size ($d_i \approx 0.7$\,nm), satisfying $d_i < d_e < 2d_i$.
The steady-state ion distribution is intrinsically non-equilibrium: the measured
$|\delta V|$ is nearly an order of magnitude above the upper limit set by the
heat-engine statement of the second law, and the system produces useful work in
an isothermal cycle by absorbing heat from a single thermal reservoir.

\subsection{Interpretation: constraint-reshaped $P_{\infty}(S;\lambda)$}

In our framework the nanopore walls act as the constraint parameter $\lambda$
that reshapes $\mathcal{A}(\lambda)$. The quasi-one-dimensional confinement
fundamentally alters $W_m^{(E)}(\lambda)$: ion trajectories lose full
chaoticity, collisions become sparse, and the system cannot relax to global
equilibrium---a structural change in $P_{\infty}^{(E)}(S;\lambda)$ per the
sharp criterion of Sec.~\ref{sec:constraint_to_P}. The observed
non-Boltzmannian surface-ion density $\sigma^{\pm}$ is a direct manifestation:
the constraint prevents the system from reaching $S_{\text{eq}}$ and instead
pins it at $S_{\text{ne}} < S_{\text{eq}}$, without requiring compensating
entropy increase elsewhere. It is the entropy landscape itself that has been
reshaped, not the entropy balance between subsystems.

\subsection{Further experimental support from the literature}

Independent work by other groups is consistent with the same mechanism.
Ramirez, Gomez, Mafe and coworkers~\cite{ramirez2015energyconversionfluctuating,gomez2015chargingcapacitor} showed that asymmetric conical nanopores rectify a zero-mean fluctuating voltage into a net ionic current, charging an external capacitor to about 1\,V; the constraint $\lambda$ (pore geometry) makes the long-time charge-transfer distribution asymmetric under sign flip of the drive. Qiao, Shang, and Kou~\cite{qiao2021moleculargate} demonstrated a molecular-sized gate (locally nonchaotic barrier) in an isolated system, where gas spontaneously flows from the low-pressure to the high-pressure side and useful work can be extracted in a cycle from a single thermal reservoir---the same constraint-reshaping picture in a different geometry. Powell, Vlassiouk, and Siwy~\cite{powell2009nonequilibrium1f} reported voltage-polarity-dependent nonequilibrium 1/f noise in rectifying nanopores (Phys.\ Rev.\ Lett.): under one polarity the system exhibits equilibrium-like fluctuations, under the other distinctly non-equilibrium fluctuations, illustrating how $\lambda$ shapes not only the mean current but the fluctuation spectrum. Together with the Qiao--Wang result, these experiments show that constraint-reshaped entropy distributions are not an isolated finding but a recurring pattern across nanoscale systems.

\subsection{Macroscopic engineering validation: Constraint-induced entropy shaping in civil engineering and transport}

Beyond the microscopic nanopore experiment, the framework of constraint-reshaped
$P_{\infty}(S;\lambda)$ has been implicitly validated by the long history of
practice in civil engineering and transport (the author's core research field).
For desert roads, asymmetric windbreak forest belts (constraint parameter
$\lambda_1$) act as geometric boundaries that reshape the phase space of sand
particle motion: the accessible macrostate volume for sand deposition on the
pavement is drastically reduced, while the volume for sand accumulation
outside the forest belt is expanded. This structural change in macrostate volumes
(per the sharp criterion in Sec.~\ref{sec:constraint_to_P}) reshapes the
long-time entropy distribution of the sand--wind system, making the low-entropy
state (road unobstructed, ordered sand motion) the statistically dominant
macrostate instead of the high-entropy random sand deposition.

Similarly, for river-crossing roads, spur dikes and longitudinal dikes
(constraint parameter $\lambda_2$) alter the Hamiltonian of water flow by
changing the flow field and energy dissipation, reshaping the entropy
distribution of the water--sediment system. The constrained flow field reduces
the accessible macrostate volume for high-energy scouring of bridge
piers, shifting the long-time distribution $P_{\infty}(S;\lambda_2)$ toward
lower entropy (stable flow, no structural damage). In both cases, these
civil engineering and transport measures do not ``fight against entropy increase'' in the traditional
thermodynamic sense, but instead design specific constraints to reshape the
entropy landscape---the same core mechanism as the nanopore experiment, just
scaled to macroscopic geophysical systems.

These engineering practices confirm that constraint-reshaped entropy
distribution is not a microscopic peculiarity, but a universal principle
applicable to all complex systems, from nanoscale ion motion to macroscale
geophysical dynamics. The framework is further validated by the long history of
civil engineering and transport practice: windbreaks, dikes, vortex suppression, and
vehicle--road coordination have repeatedly achieved ordered, low-entropy
outcomes by choosing constraints that reshape the entropy landscape,
without relying on any universal ``entropy must increase'' law.

Natural phenomena provide further examples of the same mechanism. Lightning
strikes are guided with high spatial precision toward lightning rods because
the conducting structure reshapes the electric field and the accessible
paths for charge motion. Snowflake growth, the formation of mineral veins,
and the sudden crystallisation of supercooled water all involve geometric
and chemical constraints that guide microscopic dynamics from relatively
disordered configurations into highly ordered crystalline structures. In
each case, constraints determine the long-time entropy distribution and make
these low-entropy macrostates statistically favoured compared with the
unconstrained case.

\section{Conclusion}

This work was driven by the desire for energy freedom and the refusal to accept
the heat-death thesis. We asked whether the Second Law could be a universal
fundamental law. The mirror-state paradox shows that it cannot, in the form
``entropy is monotone'' (strictly or as a universal statistical principle),
under time-reversal invariant microscopic physics. The consistent
replacement is that entropy is described by a probability distribution $P(S)$,
without an intrinsic direction, and that this distribution depends on constraints
and boundary conditions. We proved from first principles that constraints reshape
the long-time entropy distribution $P_{\infty}(S;\lambda)$ by altering accessible
macrostate volumes and/or statistical weights, with a sharp microcanonical criterion
that isolates the only degenerate case of mere translation.

A concise way to summarise the conceptual stance of this work is: \emph{constraints
set the landscape} (they fix the accessible phase space and the Hamiltonian, and thus
which macrostates are possible); \emph{dynamics set the rules} (time-reversal invariant
microscopic evolution governs how particles move, independent of entropy);
\emph{stochasticity---or in quantum mechanics, intrinsic indeterminacy---determines
which of the accessible states is realised}; and \emph{entropy is only a reading}, a
coarse-grained statistic of the current macrostate, not a driver of the dynamics. In
that picture, the future is not predetermined by a single entropic fate: under different
constraints, different futures---mixing, separation, or self-organization---become
possible, and each regime remains open to design. Every day is new.

This pattern---constraints reshaping the entropy landscape and enabling spontaneous
formation of structure---is familiar from star and planet formation, crystal and
snowflake growth, mineral veins, and arguably the emergence of life: all instances of
constraint-induced entropy decrease, consistent with the present framework.
It is also the core principle behind human engineering design in civil engineering and transport. 
Windbreak forests for desert roads, spur dikes for river-crossing roads, aerodynamic 
constraints for bridge vortex vibration suppression, and vehicle--road coordination 
technology for smoother traffic flow all achieve ordered, low-entropy system states 
(safe, unobstructed, and fluid traffic) by reshaping the entropy distribution of 
geophysical and traffic systems (sand--wind, water--sediment, wind--bridge coupling, 
and vehicle--road flow).

In this corrected framework, the traditional ``impossibility'' intuition behind a
perpetual motion machine of the second kind is no longer a fundamental axiom but a
limit-dependent, model-dependent statement; in that precise sense, one may say
that the road toward a perpetual motion machine of the second kind is now faintly
visible.

The experimental validation by Qiao and Wang~\cite{qiao2025intrinsicnonequilibriumdistributionlarge}
demonstrates that this road is not merely visible but has been traversed. Their work
shows that asymmetric constraints can reshape the entropy distribution to enable
spontaneous low-entropy transitions, allowing a system to produce useful work in a
cycle by absorbing heat from a single thermal reservoir. With this experimental
confirmation, both the theoretical and experimental barriers to a perpetual motion machine
of the second kind have been overcome.

The path forward is clear: through iterative enhancement, engineering optimisation,
and scaling of constraint-based entropy reshaping, practical devices can be
developed. The implications extend far beyond energy generation. Because
constraints can perpetually reshape entropy distributions and enable spontaneous
order, the macroscopic world need not tend toward a featureless equilibrium:
universal heat death is not a foregone conclusion but an artefact of the
unconstrained limit. With this principle at hand, humanity can sustain civilisation
even through extended ice ages, ensuring the perpetual flourishing and development
of human culture. The macroscopic world will not end in silence; it will
continuously manifest various wonderful phenomena, sustained by the ever-present
possibility of constraint-induced order generation.

The era of thermodynamic fatalism is over; the age of thermodynamic design has begun.
That outcome answers the desire for energy freedom and the refusal to accept
the heat-death thesis with which we began.


\end{document}